\documentstyle[prb,aps]{revtex}
\title{Theory of magnetoresistance in films of dilute magnetic alloys. }
 \author{L. Borda$^1$ and A. Zawadowski$^{1,2}$}
\address{$^1$Department of Theoretical Physics and Research Group of the
Hungarian Academy of Sciences,
Technical University of Budapest,\\
H-1521 Budapest, Hungary}
\address{$^2$Research Institute for Solid State Physics and Optics,
Hungarian Academy of Sciences,\\
H-1525 Budapest, Hungary}
\begin{document}
\maketitle
\draft
\begin{abstract}
Earlier a magnetic anisotropy for magnetic impurities nearby the surface of non-magnetic host was proposed in order to explain the size dependence of the Kondo effect in dilute magnetic alloys. Recently Giordano has measured the magnetoresistance of dilute Au(Fe) films for different thicknesses well above the Kondo temperature $T_K$. In this way he verified the existence of that anisotropy even for such a case where the Kondo effect is not dominating. For detailed comparison of that suggestion with experiments, the magnetic field dependence of the magnetoresistance is calculated in the lowest approximation, thus in the second order of the exchange coupling. The strength of the anisotropy is very close to earlier estimates deduced from the size dependence of the Kondo resistivity amplitude.
\end{abstract}

\pacs{PACS numbers: 72.15.Qm, 73.50 Mx, 71.70 Ej}

\section{INTRODUCTION}
\label{sec:intro}

The size dependence of the magnetoresistance observed by Giordano
\cite{label1} in thin films of dilute magnetic alloys above the Kondo
temperature may
play a decisive role in the understanding the behavior of such small
systems. In the recent decades the studies of magnetic impurities, i.e.
Fe, Cr, Mn in non-magnetic host metals (Au, Ag, Cu) have attracted
considerable interest. \cite{label2} These studies have recently been
extended to small systems by measuring their resistivity
\cite{label3,label4,label5,label6,label7,label8,label9,label10,label11,label12} and thermopower \cite{label13,label14}. The
challenging motivation of these studies has been the possibility that the
size dependence of the Kondo effect in thin films and wires might give
some information on the size of the spin compensation cloud formed by the
conduction electrons, which is responsible for the formation of the
magnetic singlet ground state \cite{label15}. That possibility is
ruled out as the Kondo impurity experiences only the level spacing of the
conduction electrons at the impurity site \cite{label16,label17}, thus
there is no such effect except of very small granular samples. Furthermore,
there are experiments where no size dependence occurs \cite{label11}.
Recently, two attempts have been made to explain the size dependence: (i)
static disorder proposed by Martin, Wan and Phillips
\cite{label18} (ii) surface spin anisotropy for magnetic impurities due to
spin-orbit interaction between the nonmagnetic host atoms and electrons.
\cite{label17,label19,label20,label21}. These two suggestions can be applied in
the opposite limits, namely in the dirty and ballistic regions.

 If the normal vector of the surface is ${\bf n}$, then the anisotropy
energy $H_a$ is 
\begin{equation}
H_a=K_d ({\bf n}{\bf S})^2,
\label{aniz}
\end{equation}
where ${\bf S}$ is the spin operator of the impurity spin and $K_d$ is the
strength of the anisotropy which depends on the distance, $d$ of the impurity
measured from the surface. The suggested spin anisotropy nearby the
surface hinders the spin dynamics, thus it can affect the amplitude of the
Kondo effect, but for integer and half-integer impurity spins in different
ways \cite{label19,label22}. The occurance of that anisotropy is
independent of the Kondo effect as it can be obtained by straightforward
perturbation theory without any logarithmic corrections. At the same time
the evidences are growing for the size dependence but only in the temperature
region governed by the Kondo effect \cite{label10,label14}.

In order to clarify the relevance of the magnetic
surface
anisotropy Giordano \cite{label1} measured the magnetoresistance of thin
gold films of thickness $410\text{{ }\AA { }}$ and $625\text{{ }\AA { }}$ with $30 ppm$
 Fe impurities.
At $1.4 K$, above the Kondo temperature $T_K=0.3 K$ he found an
essential difference in the magnetic field dependence of the resistivity
by measuring these
two samples. That challenging phenomena was attributed by him to the surface
anisotropy.

The basic idea is that due to the strong
spin-orbit scattering of the conduction electrons on the  non-magnetic
host atoms a magnetic anisotropy is developed for the magnetic ions nearby
the surface. The magnetic scattering (exchange) of the conduction
electrons by the impurity ion has a strong angular dependence as the
scattered waves are d-type, which are hybridized with the localized
magnetic d-orbitals of the impurity. The scattered waves traveling
through the samples suffer spin-dependent scatterings due to the
spin-orbit scattering. As far as the host atom scatterers are in the
ballistic region measured from the magnetic ion, the scattering
amplitudes for spin-flip and spin-conserving scatterings depend on the
angular momenta of the conduction electron wave functions. In this way the
magnetic impurity receives information about the positions of the host
atoms, 
thus the
geometry of the sample in the ballistic region around the host atom. The
phase information is lost outside the ballistic region. That phenomenon
results in forming of a surface magnetic anisotropy, $K_d$ which 
depends inversely on the distance $d$ measured from the surface, $K_d
\sim d^{-1}$ and
exhibits the anisotropy axis perpendicular to the surface. The anisotropic
energy is $K_d (S^z)^2$ where $z$ is the normal direction of the surface, and $S^z$ is the $z$-component of the spin.
Recently that results has been generalized by Fomin {\it et.al.}
\cite{label21} for arbitrary geometry of the
sample and the actual calculation has been performed for a sample of slat
shape, also. Concerning the boundary conditions they have shown that a 
moderate roughness of the surface even may enhance the anisotropy.

In the magnetoresistance experiments on bulk samples 
as the magnetic field increases the
level spacing of magnetic atomic levels become larger thus as the magnetic
energy exceeds the temperature the spin-flip electron scattering is
gradually frozen out and only the lowest level is occupied. That effect is very sensitive on the actual level
spacing
of the atomic levels as a function of the magnetic field. Considering a
magnetic atom in the neighborhood of the surface an essential level
splitting occurs even without magnetic field
\cite{label17,label19,label20}. The magnetic field splits these levels
further.
The magnetic field saturating the magnetoresistance for a given temperature
can be much larger as the energy of the lowest level only occupied in high field is shifted
upwards by the surface anisotropy relatively to the others, see Fig. \ref{fig0}.

Carrying out that calculation a conceptional problem arises. In the derivation of the surface anisotropy the Anderson model had to be used with nonzero angular momentum of the localized $d$-level. Only in case of spin $S=5/2$ the orbital momentum is quenched, thus for $S\neq 5/2$ a more general Hamiltonian must be used as it is discussed e.g. by Nozi\`eres and Blandin\cite{footnote}  but that is actually a hopeless task, because of the large number of terms and couplings. Here we follow the simplification applied in Ref.\cite{label19} and \cite{label20}, where we kept a simple spin Hamiltonian with only one orbital channel for the conduction electrons but the anisotropy obtained for $S=5/2$ is generalized for arbitrary spin. There is no reason to believe that the saturation cannot be demonstrated in that simplified model even if details and the value of the Kondo exchange coupling value are somewhat changed.

The present paper is devoted to calculate the magnetoresistance of thin
films taking into account the magnetic surface anisotropy in the ballistic
region for integer and half-integer impurity spins. Fitting the data obtained from Au(Fe) films by Giordano \cite{label1} the strength of the surface
anisotropy is determined. That anisotropy strength is compared with the
value estimated from the size dependence of the Kondo resistivity
\cite{label19,label20}. They are found in the same order of magnitude and
their difference which is about a factor four can be easily accounted for 
the difference in the mean free path due to the different sample preparations or to the different surface roughness\cite{label21}.

Finally it should be emphasized that the calculation is carried out in the lowest nonvanishing order of perturbation theory ignoring any corrections of higher order of Kondo type which is justified if the temperature $T$ is much higher than the Kondo temperature. In the case of Giordano's experiment\cite{label1} $T=1.4 K$ and $T_K=0.3 K$. In that case the measured resistivity shows logarithmic dependence and size effect as well. The size dependence of the magnetoresistance is twice as strong as in case of the logarithmic part of the resistivity. (See Ref. \cite{label1} Fig. 4(b).) That indicates that the size dependence of the magnetoresistance cannot be explained simply by the size dependence of the logarithmic coefficient $\tilde{B}$ of the resistivity replacing $\tilde{B}\log_{10}T$ by $\tilde{B}\log_{10}(T^2+{\bf B}^2/\alpha^2)^{1/2}$ where $\alpha$ is in the order of unity and ${\bf B}$ is the magnetic field\cite{plus1}. The magnetoresistance for infinite system was calculated many years ago by Abrikosov in the leading logarithmic approximation\cite{abrikosov}, where $\alpha$ cannot be determined. As far as we know no major improvement has been achieved since that. Thus a calculation, where the surface anisotropy and the Kondo term in the next to leading logarithmic approximation as in Ref.
\cite{label20} are taken into account is not feasible. Therefore, the goal of the present paper is only to demonstrate the strong size dependence even without Kondo correction, therefore the comparison with experiment must be taken with care.

The paper is organized as follows. The model is given in Sec. \ref{sec:model}. In Sec. \ref{sec:split} the energy
level splitting for the magnetic ions due to the magnetic field is calculated
in the presence of surface anisotropy for different directions of the magnetic
field relative to the surface. In Sec.\ref{sec:3} the Boltzmann equation is
solved in a simple relaxation time approximation for a uniform anisotropy. In
Sec.\ref{sec:4} the magnetoresistance is calculated for a realistic thin
film where the anisotropy depends on the positions of the impurities and an
average is taken over the positions. In Sec.\ref{sec:5} the fit for
experimental data is presented. In Sec.\ref{sec:6} the results are
briefly
discussed. In the Appendix some details of the calculations are
presented.
\section{The model}
\label{sec:model}
The model used in the present paper is based on the Hamiltonian
\begin{equation}
H=H_a+H_e+H_{int.} .
\label{3h}
\end{equation}
The anisotropy Hamiltonian is
\begin{equation}
H_a=\sum\limits_i \!\!( K_{d_i}({\bf n}{\bf S_i})^2+
\mu \hbar g{\bf BS_i}), 
\label{ha}
\end{equation}
with the anisotropy constant $K_{d_i}$ where $i$ labels the magnetic 
impurity with spin operator ${\bf S_i}$, ${\bf B}$ is the magnetic field
vector, $\mu $ is the Bohr magneton and $g$ is the gyro-magnetic factor
($g=2$).

The electron Hamiltonian $H_e$ describes the free electrons
\begin{equation}
H_e=\sum\limits_{{\bf k},\sigma }\!\! {{\hbar ^2k^2}\over{2m_e}} 
a^+_{{\bf k},\sigma } a_{{\bf k},\sigma } +
B\mu \sum\limits_{{\bf k},\sigma }\!\! \sigma ^z a^+_{{\bf k},\sigma }
a_{{\bf k},\sigma } ,
\label{he}
\end{equation}
where $m_e$ is the electron mass, ${\bf k}$ is the wave number of the
electron created and annihilated with spin $\sigma $ by the operators
$a^+_{{\bf k},\sigma}$ and $a_{{\bf k},\sigma}$, respectively. In order
to avoid the difficulties of the boundary condition on the surface the 
electron states are extended to the unlimited space. In Ref. \cite{label21} it is
shown that the changing the boundary conditions does not affect the main
features of the anisotropy.

The interaction Hamiltonian is the usual s-d Hamiltonian
\begin{equation}
H_{int.}=-\sum\limits_{{\bf k},{\bf q},i}\!\!
J_{{\bf kq}}e^{i({\bf k}-{\bf q}){\bf R_i}}\biggl[ 
\sum\limits_{\sigma=+,-}\!\!
S_i^za^+_{{\bf q},\sigma}a_{{\bf k},\sigma}+a^+_{{\bf q},-}a_{{\bf k},+}
S^+_i +a^+_{{\bf q},+}a_{{\bf k},-}S^-_i \biggr] ,
\label{hint}
\end{equation}
where $J_{{\bf kq}}$ is the momentum dependent exchange coupling  and
${\bf R_i}$ is the position of $i$th impurity. In principle for magnetic
impurities with localized d-electrons the momentum dependence is important,
but just to simplify the calculation as it is discussed at the end of the Introduction it will be dropped, thus $J_{{\bf kq}}
\rightarrow J$. Such simplification does not modify the magnetic transport
itself in an essential way but that is certainly inadequate in the derivation 
of the anisotropy constant itself \cite{label19}.

\section{SPLITTINGS OF THE ENERGY LEVELS OF THE IMPURITY SPIN DUE TO THE
ANISOTROPY AND MAGNETIC FIELD}
\label{sec:split}

The Hamiltonian of the single impurity nearby the surface with normal vector
${\bf n}$ in a magnetic field ${\bf B}$ is
\begin{equation}
H_a=K_d(${\bf n}{\bf S}$)^2+\mu g\hbar {\bf B}{\bf S},
\label{anizmagn}
\end{equation}
where $K_d$  is the anisotropy constant, ${\bf S}$ is the spin operator of
the impurity. In general the two terms of Hamiltonian (\ref{anizmagn}) cannot
be diagonalized separately except the magnetic field ${\bf B}$ is
perpendicular to the surface, as in the general case the two terms do not
commute. Therefore, three different cases will be treated
\begin{description}
\item[(i)] {\bf B} $\parallel $ {\bf n}
\item[(ii)] {\bf B} $\perp$ {\bf n}
\item[(iii)] arbitrary angle between {\bf B} and {\bf n}.
\end{description}

The analytical results can be given for case (i) for arbitrary spin $S$,
the case (ii) can be solved analytically and the results are presented for 
$S=2$ in the Appendix. The general case will be demonstrated only for $S=1$.

{\it case(i)}

In this case the Hamiltonian (\ref{anizmagn}) is
\begin{equation}
H=K_d(S^z)^2+\mu \hbar gBS^z .
\label{anizmagnz}
\end{equation}
Introducing the $|m\rangle $ eigenvectors of the spin operator $S^z$
labelled by the
azimuthal momentum $m$ as
\begin{equation}
S^z|m\rangle=\hbar m|m\rangle ,
\label{Szsajat}
\end{equation}
where $-S\leq m\leq S$, the energy eigenvalues of Hamiltonian
(\ref{anizmagnz}) are
\begin{equation}
E_m=\hbar ^2m^2K_d+\mu \hbar gmB .
\label{eigenvalsz}
\end{equation}

{\it case(ii)}

The Hamiltonian for ${\bf n}||z$ and ${\bf B}||x$ is
\begin{equation}
H=K_d(S^z)^2+\mu \hbar gBS^x.
\label{anizmagnx}
\end{equation}
The eigenvalues of this Hamiltonian for $S=2$ are shown in Fig. \ref{fig1}. As the two terms of Hamiltonian
(\ref{anizmagnx})
are not commuting there are no level crossing indicating that the eigenvectors
are given by very complicated expressions not given here. The analytical expressions of the eigenvalues are given in the Appendix.

{\it case(iii)}

In order to demonstrate the level crossing we present the result for $S=1$
and an arbitrary magnetic field ${\bf B}=(B_x,0,B_z)$.
The eigenvalues are given in the Appendix and shown in Fig. \ref{fig2}.
If $B_x=0$, then the levels are crossing similarly shown in Fig. \ref{fig2}.  by dashed lines. The exact eigenvalues for
${B_x\over B_z}=0.1$ are shown in the figure by solid lines.

For ${B_x\over B_z}=0.1$ the levels are close to those of case $B_z=B,B_x=0$ 
but the
weak perpendicular field splits the degeneracy at $\mu \hbar gB=\hbar ^2 K_d$.
As $B_x$ increases the level hybridization can be so large that the
crossings are hard to be recognized. At low fields the spin of the
eigenstates are directed by the anisotropy while in high magnetic field by
the field.

\section{SOLUTION OF THE BOLTZMANN EQUATION FOR FIXED VALUE OF THE
ANISOTROPY CONSTANT}
\label{sec:3}
In this section we solve the Boltzmann equation in external magnetic
field for the electrons and
calculate the electrical resistivity for a fixed strength of the
anisotropy constant for all of the impurities. We calculate the
contribution of the impurities in the magnetoresistance. In this
calculation we neglect the magnetic field term in the drift part of the
Boltzmann equation.

We use the relaxation time approximation and we ignore the momentum
dependence of the exchange coupling, thus isotropic scattering on the
impurity is assumed. The correct exchange coupling contains a factor
$P_{l=2}(cos\vartheta_{{\bf k},{\bf k'}})$ for the scattering on a magnetic
impurity with d-level, where $P_l$ is the Legendre polynom and
$\vartheta_{{\bf k},{\bf k'}}$ is the angle between the momenta of the
incoming
and scattered electrons. That may change some details of the result, but
certainly does not affect the overall dependence on the magnetic filed.

Following closely the work of Yoshida \cite{yoshida} the Boltzmann equation linearized in the
electronic field ${\bf E}=({\it E},0,0)$ takes the form
\begin {equation}
{{{\partial f_0(E_k)}\over{\partial E_k}}} {{\hbar k_x}\over {m_e}} e{\it E}+
{{({{\partial f^{\pm}}\over {\partial t}})}}_{coll.}=0 ,
\label{bolt1}
\end {equation}
where $f_0$ is the equilibrium distribution function depending on the
energy $E_{\bf k}$ of electron, $f^{\pm}$ are the nonequlibrium
distribution function for spin parallel and anti-parallel to the
magnetic field. The collision term is denoted by $( )_{coll.}$ . For the
electrons free electron approximation is applied with uniform mass $m_e$
and $(k_x,k_y,k_z)$ denote the momentum of the electron. The electronic
charge is $-e$ and the volume of the sample is $V$.

The solution of Eq. (\ref{bolt1}) can be found in form of the following
Ansatz
\begin{equation}
f^{\pm} = f_0(E_k \pm {1\over {2}} \mu \hbar gB)-k_x{\it E}\Phi ^{\pm}(E_{k,\pm })
{{{\partial f_0(E_k \pm {1\over {2}} \mu \hbar gB)}\over{{\partial E_k}}}}=
f_0-\Delta f^{\pm} ,
\label{ansatz}
\end{equation}
where the functions $\Phi ^{\pm}$ should be determined. The collision part
of the Boltzmann equation is
\begin{eqnarray}
\left({{{\partial f^{\pm}}\over{\partial t}}}\right)_{coll.}&=&
{1\over V}\sum\limits_{{\bf k}'}\!\! W({\bf k}'\pm\rightarrow {\bf k}\pm)
f^{\pm}({\bf k}')(1-f^{\pm}({\bf k}))- \nonumber\\
&-&W({\bf k}\pm\rightarrow {\bf k}'\pm)
f^{\pm}({\bf k})(1-f^{\pm}({\bf k}'))+\nonumber\\
&+& W({\bf k}'\mp\rightarrow {\bf k}\pm)
f^{\mp}({\bf k}')(1-f^{\pm}({\bf k}))-\nonumber\\
&-&W({\bf k}\pm\rightarrow {\bf k}'\mp)
f^{\pm}({\bf k})(1-f^{\mp}({\bf k}')) ,
\label{collterm}
\end{eqnarray}
where $W({\bf k}\pm\rightarrow {\bf k}'\pm)$ is the transition probability of
an electron
with momentum ${\bf k}$ scattered into the state with ${\bf k}'$, the
spins are denoted by $+$ and $-$. That probability does not contain the
occupation numbers for the incoming and outgoing electrons, but that
contains the average over the initial states for the localized spins.

The current can be expressed in term of the functions $\Phi ^{\pm}$ in the
usual way for electron spin $+$ and $-$ as
\begin{equation}
j=j^+ +j^- ,
\label{j}
\end{equation}
and
\begin{eqnarray}
j^{\pm}&=&-e{1\over V}\sum\limits_k\!\! {{\hbar k_x}\over{m_e}}\left[{
f^{\pm}(E_{k\pm})-f_0(E_{k\pm})}\right] \nonumber\\
&=&{{e\hbar}\over{m_e}}\int dk\int d\Omega_k k^2k_x^2{\it E}\Phi^{\pm}
(E_{k}\pm\mu\hbar B){{\partial f_0(E_{k}\pm\mu\hbar B)}\over{\partial E_k}}
\nonumber\\
&=&{\it E}{{e\hbar}\over{m_e}}\left[{{{2m_e}\over{\hbar ^2}}}\right]^{3\over 2}
{1\over{6\pi ^2}}\int dEE^{3\over 2}\Phi^{\pm}
(E_{k}\pm\mu\hbar B){{\partial f_0(E_{k}\pm\mu\hbar B)}\over{\partial
E_k}} ,   
\label{jpm}
\end{eqnarray}
where $d\Omega_k$ is the solid angle for the electron with momentum ${\bf k}$.

The conductivity is from Eqs.(\ref{j}) and (\ref{jpm}) is
\begin{eqnarray}
\sigma&=&{{e\hbar}\over{m_e}}\left[{{{2m_e}\over{\hbar ^2}}}\right]^{3\over 2}
{1\over{6\pi ^2}}\int dEE^{3\over 2}\times \nonumber\\
&\times &\left[{
\Phi^{+}
(E_{k}+\mu\hbar B){{\partial f_0(E_{k}+\mu\hbar B)}\over{\partial E_k}}+
\Phi^{-}
(E_{k}-\mu\hbar B){{\partial f_0(E_{k}-\mu\hbar B)}\over{\partial E_k}}}
\right].
\label{cond}
\end{eqnarray}

The next task is to calculate the transition probability $W$. In order 
to describe the magnetic splitting of the conduction electrons it is 
adequate to use a coordinate system given by unit vectors $\tilde x,
\tilde y, \tilde z$ with
$\tilde z$-direction parallel to the field (see Fig. \ref{fig3}.). 
In this case we have to
introduce three set of states for the impurity spin. In Sec. \ref{sec:split}. we used
the system attached to the anisotropy axis thus it is in the frame of
the sample, where the spin states are organized in the column vector

\begin{equation}
\left[{
\begin{array}{c}
|S\rangle \nonumber\\
... \nonumber\\
|m\rangle \nonumber\\
... \nonumber\\
|-S\rangle
\nonumber
\end{array}}
\right] ,
\label{smatr}
\end{equation}
with the magnetic quantum number $m$, $-S \leq m\leq S$. The exact
energy eigenstates in the crystalline field combined with the magnetic
field determined in Sec. \ref{sec:split}. form 
the vector

\begin{equation}
\left[{
\begin{array}{c}
|1\rangle \nonumber\\
... \nonumber\\
|r\rangle \nonumber\\
... \nonumber\\
|2S+1\rangle \nonumber
\end{array}}
\right] ,
\label{evmatr}
\end{equation}
finally the states quantized according to the magnetic field $B$ are 
denoted as

\begin{equation}
\left[{
\begin{array}{c}
|\tilde S\rangle \nonumber\\
... \nonumber\\
|\tilde m\rangle \nonumber\\
... \nonumber\\
|-\tilde S\rangle \nonumber
\end{array}}
\right] .
\label{stmatr}
\end{equation}
An arbitrary operator $A$ can be given in terms of the exact energy states 
$|r\rangle$ expressed by the states $|\tilde m\rangle$ as

\begin{eqnarray}
A&=&\sum\limits_{r,r'}\!\! |r\rangle\langle{r}|A|r'\rangle\langle{r'}|
\nonumber\\
&=&\sum\limits_{r,r',m,m'}\!\!\ |r\rangle \langle{r}|\tilde m\rangle 
\langle{\tilde m}|A|\tilde m'\rangle\langle{\tilde m'}|
r'\rangle\langle{r'}|\nonumber\\
&=&\sum\limits_{r,r'}\!\!\ |r\rangle A_{r,r'}\langle{r'}|,  
\label{aope}
\end{eqnarray}
where $A_{r,r'}$ is a matrix.

For the representation of the exchange Hamiltonian an arbitrary axis can
be used and as it mentioned before, the direction of magnetic field is the
most adequate one, thus ${\bf B}\parallel {\bf{\tilde z}}$. In that system
the 
impurity spin operators are $\tilde {S^x},\tilde {S^y},\tilde {S^z}$ and
$\tilde {S^\pm}=\tilde {S^x}\pm i\tilde {S^y}$, for which

\begin{eqnarray}
\langle{\tilde m}|\tilde {S^z}|{\tilde m'}
\rangle
&=&\tilde m\delta_{\tilde m,
\tilde m'} \nonumber \\
\langle{\tilde m}|\tilde {S^+}|{\tilde m'}
\rangle
&=&\sqrt{S(S+1)-\tilde m
(\tilde m +1)}\delta_{\tilde m,\tilde m'+1} \nonumber \\
\langle{\tilde m}|\tilde {S^-}|{\tilde m'}
\rangle
&=&\sqrt{S(S+1)-\tilde m
(\tilde m-1)}\delta_{\tilde m,\tilde m'-1} 
\label{sope}
\end{eqnarray}
hold.

In order to use Eq. (\ref{aope}) the matrix element 
$\langle r|\tilde m\rangle $ must be known. However, in Sec. II. we
expressed
the states $|r\rangle$ with anisotropy in terms of states $|m\rangle$
given in the frame of the sample. Thus
\begin{equation}
\langle r|\tilde m\rangle =
\sum\limits_m\!\! \langle r|m\rangle \langle m|\tilde m \rangle .
\label{rmmatr}
\end{equation}
The matrix elements $\langle r|m \rangle$ are calculated in Sec. \ref{sec:split}.,
while $\langle m|\tilde m\rangle$ follows from a simple rotation by 
the angle between the directions given by the sample and magnetic field.
In this way the matrix element $\langle r|\tilde S^{\pm}|r'\rangle$ and
$\langle r|\tilde S^z|r'\rangle$ can be calculated directly by
applying Eq. (\ref{aope}) and the details of the calculations are
straightforward and, therefore, those are not presented.

In this coordinate system the interacting Hamiltonian is
\begin{equation}
H_{in}=-J\sum\limits_{i,\alpha,\beta}\!\!
\tilde{\bf S_i}\Psi^+_{\alpha}({\bf R_i})\tilde{\bf {\sigma}}_{\alpha 
\beta}
\Psi_{\beta}({\bf R_i}) ,
\label{inte}
\end{equation}
where $\Psi$ is the electron field operator at the impurity site
${\bf R_i}$
The transition probability
\begin{equation}
W(k\tilde {\sigma}\rightarrow k'\tilde {\sigma}')={2\pi \over{\hbar}} J^2 \sum
\limits_r\!\! p_r|\langle{r,\tilde{\sigma}}|\tilde {\bf S} \tilde {\bf
{\sigma}}|{r',
\tilde{\sigma}'}
\rangle
|^2 \delta (E_{k\tilde {\sigma}}+
E_r-E_{k'\tilde {\sigma'}}-E_{r'}) ,
\label{tranprob}
\end{equation}
where the state $|r,\sigma\rangle$ is the product of the impurity state
$|r\rangle$ and the spin state $|\tilde{\sigma}\rangle$ which is quantized
along the magnetic field.
The occupation number for the exact eigenstate ($|r\rangle $) is
\begin{equation}
p_r={{e^{-\beta E_r}}\over{\sum\limits_r\!\! 
e^{-\beta E_r}}} ,
\end{equation}
and $E_{k\tilde {\sigma}}=E_k+\mu \hbar B\tilde {\sigma}^z$.
In the following the three cases introduced in Sec. \ref{sec:split}. are discussed with
the normal vector ${\bf n}=\tilde{\bf z}$.
The calculation is performed in case(i) for arbitrary spin, in case(ii) for
$S=1,{3\over2}$ and in case(iii) only for $S=1$. The cases (ii) and (iii) are
discussed only for the $S=1$ because of the
difficulties in the
analytical calculation.

{\it case(i)}

Using Eq.\ (\ref{tranprob}) and Eq.\ (\ref{collterm}) the collision terms can
be given as
\begin{eqnarray}
\left({
{{\partial f^{\pm}}\over{\partial t}}}
\right)_{no flip}&=&
{1\over V}\sum\limits_{{\bf k}'}\!\! {{2\pi}\over{\hbar}}J^2
\sum\limits_m\!\! p_mm^2{\hbar}^2\delta (E_{{\bf k}'}-E_{\bf k})
f^{\pm}(E_{{\bf k}'})(1-f^{\pm}(E_{{\bf k}}))- \nonumber\\
&-&{1\over V}\sum\limits_{{\bf k}'}\!\! {{2\pi}\over{\hbar}}J^2
\sum\limits_m\!\! p_mm^2{\hbar}^2\delta (E_{{\bf k}}-E_{{\bf k}'})
f^{\pm}(E_{{\bf k}})(1-f^{\pm}(E_{{\bf k}'})) ,
\label{colltermz1}
\end{eqnarray}
\begin{eqnarray}
\left({
{{\partial f^{\pm}}\over{\partial t}}}
\right)_{flip}&=&
{1\over V}\sum\limits_{{\bf k}'}\!\! {{2\pi}\over{\hbar}}J^2
\sum\limits_m\!\! p_m{\hbar}^2(S(S+1)-m(m\mp 1))\delta (E_{{\bf k}'}-
E_{\bf k}-K_d{\hbar}^2(1\mp 2m))
\nonumber\\
&\times &f^{\pm}(E_{{\bf k}'})(1-f^{\pm}(E_{{\bf k}}))- \nonumber\\
&-&{1\over V}\sum\limits_{{\bf k}'}\!\! {{2\pi}\over{\hbar}}J^2
\sum\limits_m\!\! p_m{\hbar}^2(S(S+1)-m(m\pm 1))
\delta (E_{{\bf k}}-E_{{\bf k}'}-K_d{\hbar}^2(1\pm 2m))\nonumber\\
&\times &f^{\pm}(E_{{\bf k}})(1-f^{\pm}(E_{{\bf k}'})) ,\nonumber\\ 
\left({
{{\partial f^{\pm}}\over{\partial t}}}
\right)_{coll.}&=&\left({
{{\partial f^{\pm}}\over{\partial t}}}
\right)_{flip}+\left({
{{\partial f^{\pm}}\over{\partial t}}}
\right)_{no flip} ,
\label{colltermz}
\end{eqnarray}
where in the first term the inscatterings drop out due to the
Ansatz (\ref{ansatz}). After linearization in $\Delta f$ and using
\begin{equation}
{1\over V}\sum\limits_{{\bf k}'} \rightarrow
\int {{{\bf d^3k}'}\over{(2\pi )^3}} \rightarrow
\int {{d\Omega _{k'}}\over{4\pi}}\int{{\rho}(E_{k'})dE_{k'}} \rightarrow
{{m_ek_F}\over{2\pi ^2}}\int {{d\Omega _{k'}}\over{4\pi}}\int dE_{k'} ,
\end{equation}
the collision term takes the form
\begin{eqnarray}
\left({
{{\partial f^{\pm}}\over{\partial t}}}
\right)_{coll.}&=&
{{m_ek_F\hbar}\over{\pi}}J^2\sum\limits_m\!\!
p_m\hbar ^2\biggl [
m^2\Delta f^{\pm}(E_{k\pm})+\nonumber\\
&+&(S(S+1)-m(m\mp 1))\times 
\left({f_0(E_{k\mp}+(1\mp 2m)K_d \hbar ^2)\Delta f^{\pm}(E_{k\pm})}\right)
\nonumber\\
&-&(S(S+1)-m(m\pm 1))\times
\left({f_0(E_{k\mp}+(1\mp 2m)K_d \hbar ^2)\Delta
f^{\pm}(E_{k\pm})}\right)
\biggr ].
\label{III21}
\end{eqnarray}
Using the Ansatz (\ref{ansatz}) it takes the form
\begin{equation}
\left({
{{\partial f^{\pm}}\over{\partial t}}} \right)_{coll.}=
2\pi {\rho}(E_F)J^2k_x{\it E}\Phi ^{\pm}(E_{k\pm})
{{\partial f_0(E_{k\pm})}\over{\partial E_{k}}} F^{\pm}(E_{k\pm}),
\label{III22}
\end{equation}
where
\begin{eqnarray}
F^{\pm}&=&\sum\limits_m\!\!
p_m\biggl [
m^2+(S(S+1)-m(m\mp 1))\biggl (
f_0(E_{k\mp}+(1\mp 2m)K_d\hbar ^2)+\nonumber\\
&+&e^{-\beta (K_d\hbar ^2(\mp 2m+1)\mp
2\mu \hbar B)}(1-f_0(E_{k\mp}+(1\mp 2m)K_d\hbar ^2))\biggr ) \biggr ].
\label{fpm}
\end{eqnarray}
Then from the Boltzmann equation (\ref{bolt1}) follows that
\begin{equation}
\Phi ^{\pm}(E_{k\pm})=
-{e\over{2\pi {\rho}(E_F)m_eJ^2}}{1\over{F^{\pm}(E_{k\pm})}} .
\label{phipmz}
\end{equation}

{\it case(ii)} and {\it case(iii)}

The calculation of magnetoresistance in {\it case(ii)} and
{\it case(iii)} is similar to the one presented above.  
\section{Calculation of magnetoresistance for realistic films}
\label{sec:4}
In the calculation presented in the previous section the strength of the
anisotropy is uniform, thus no realistic dependence of the magnetoresistance
on the thickness of the film is obtained. In this section we calculate the
magnetoresistance for thin films and take the position dependence of the
anisotropy factor $K_d$ into account.

In Fig. \ref{fig5}. the conductivity for $S=2$ is shown as the
function of external magnetic field for a fixed $K_0$ value of the
anisotropy constant. At ${{\mu \hbar gB}\over{K_0\hbar ^2}}\simeq 4$ there is a
local minimum in the conductivity curve due to the level-crossing shown in
Fig. \ref{fig2}.

In the curves measured by Giordano \cite{label1} there is no such a minimum
because it depends on the specific value of anisotropy strength $K_d$.
That value depends on the
distance of magnetic impurity measured from the surface of the film. Thus,
if
we take an average over the positions of the impurities the minima disappear and the
conductivity will be a smooth function of magnetic field as shown in
Fig. \ref{fig6}. In the case of a film it can be assumed, that both
surfaces
contribute to the anisotropy in a similar way, thus the impurity in a
distance $d$ from one of the surfaces (as shown in Fig. \ref{fig3}.)
experiences the anisotropy
\begin{equation}
K_d(t)= {{\alpha}\over{d}}+{{\alpha}\over{t-d}}.
\label{aniztd}
\end{equation}
The average can be taken as follows. If the concentration of the impurities with randomly distributed anisotropy is
uniform in the sample we only have to take an average over the values of
$d$ and it means an integral, but we can approximate this
average with a sum. But there is a difficulty: considering the thickness
of the film compared with the electron mean free path $l_{el}$,there are
two different limits
(i) $t < l_{el}$ or (ii) $t>>l_{el}$ and it is not trivial that the final
expression of resistance does not depend on which limit are taken. In
Ref.\cite{label19} the
calculations carried out in these two limits. In the first case the
electrical
resistivity is calculated by averaging over the inverse electron lifetime, 
while in the second case
the sample is considered as a set of parallel resistors of equal size,
where each
resistor represents a stripe in the sample with different constant $K$. 
There it has been shown by writing the resistivity as a sum of the host 
resistivity $\rho _{host}$ and the contribution of the impurities 
$\rho _{imp}$, $\rho =\rho _{host}+\rho _{imp}$ that in the limit $\rho _{imp}
<< \rho _{host}$ in both cases the final expressions take the same  form:
\begin{equation}
\rho (t,T)={1\over t}\int\limits_0^t \rho \left[{
K(x,t),T}\right] dx ,
\label{roatlag}
\end{equation}
where $t$ is the thickness of the sample and $T$ is the temperature. This
calculation of the average is
valid for finite magnetic field, as well. In the actual calculation the integral in Eq.(\ref{roatlag}) is replaced by a sum of parallel resistor terms with different anisotropy constant $K$. The final result for $S=2$ is shown in Fig. \ref{fig6}. for different temperatures. The local minimum due to level crossing disappears as it is expected. The calculated conductivity vs magnetic field curves can be seen for $S=1$ in Fig. \ref{fig4}. for different directions of magnetic field.

\section{Comparison with experiments}
\label{sec:5}
The size dependence of the magnetoresistance in thin films have been
observed by Giordano \cite{label1} by measuring $Au$ films with about
$30 ppm$ $Fe$ and the thickness of films were $410$ and $625$ \AA { }
. The magnetoresistance measurements were performed at $T=1.4 K$, well 
above the Kondo temperature $T_K =0.3K$ and the external magnetic field was
perpendicular and parallel to the surface of the films.
In this paragraph the theoretical curves
calculated in the previous section will be compared with those
experimental
data. As it is shown in Fig. 6. of Ref.
\cite{label1} there was no significant difference between experimental curves
of two different directions of the external magnetic field. In
the theoretical
curves there is about 10 percent difference between the two cases
at intermediate $(\mu \hbar gB \simeq 4 K_d\hbar ^2)$ values of magnetic field,
as there can be seen in Fig. \ref{fig4}. for $S=1$, but we do not have theoretical curve for the realistic spin value $S=2$.
The absence of the difference in the measured curves might be  also due to
the surface roughness. In that case the
direction of
anisotropy for the magnetic impurities may be different for impurities
placed at
different positions in the sample.
 Therefore, the measured value of magnetoresistance will be the value
averaged over the distribution of the angle between the surface and
the magnetic field.

For fitting of the experiments the theory without logarithmic corrections is carried out, thus that can indicate the strong size dependence, but the parameters obtained can be modified if the logarithmic corrections could be also taken into account.

 In the fitting for the two curves (for the films with thickness 625 and 410
\AA { }) the same value of anisotropy strength, $\alpha$ was taken. In the
fitting procedure only the anisotropy strength, $\alpha$ and the coupling
$J$ is 
determined,
and there are no other free parameters in the calculation presented previous
paragraphs. The value of $\alpha$ is determined by the scale on the $B$
axis and
$J$ is by the amplitude of $\Delta \rho$. Actually, the fitting was
carried out for the
sample with thickness 625 \AA { } and the obtained parameters were used for the
other one except the slight change in the concentration. If the concentration
of Fe in the film with size 625 \AA { } is $c$, then the other contains $0.9 c$ Fe.

 The fitted curves is shown in Fig. \ref{fig7}. As there can be seen in
Fig. \ref{fig7}., the calculated curves of resistivity is in excellent
agreement with the measured ones even without logarithmic corrections. The fitted value of anisotropy strength
$\alpha$ is
\begin{equation}
{{\alpha}\over{k_B}}=42 \text{ K\AA { }} .
\label{fittedalpha}
\end{equation}
As the logarithmic terms (see in Ref.\cite{label1} Fig. 4(b)) should have less size dependence, therefore counting only non-logarithmic effects, the strength $\alpha$ may be somewhat underestimated.
In the calculation the constant $\sigma _0$ was introduced as the unit of conductivity 
(in $\hbar =1$ units) as
\begin{equation}
\sigma _0={{\rho ^2(\varepsilon _F)}\over {J^2}} {{8\pi ^3}\over 12}
{{m_e^5}\over {e^2}}.
\label{sigma0}
\end{equation}
At the fitting procedure we use the measured value of $\sigma _0$, 
${{\sigma _0}^{-1}}\simeq 0.6 n\Omega cm$.

\section{Discussion}
\label{sec:6}
Until recently the surface anisotropy for magnetic impurities in metallic
hosts with strong spin-orbit scattering has been deduced from the reduction 
in the Kondo resistivity amplitude measured by Giordano and his co-workers
\cite{label3,label4,label5,label6,label7,label8,label9}. The anisotropy is independent of
the Kondo effect according to the theories \cite{label19,label20,label21}.
Giordano \cite{label1} demonstrated the size effect in the magnetoresistance
where the level splittings modify the saturation by the magnetic field which can result in the stronger size dependence compared to the Kondo term.
In the present paper we gave the detailed theory of magnetoresistance which
supports the suggestions of Giordano \cite{label1}. The strength of the 
anisotropy $\alpha = 42$ K\AA { } $k_B$ (see Eq. \ref{fittedalpha} ) is obtained  by fitting the
Au(Fe) samples of two different thicknesses and both fits give the same value.
That constant have been also deduced from the size dependence of Kondo anomaly
for the same compounds \cite{label20}. The value obtained there is somewhat 
different $\alpha =247$ K\AA { } $k_B$. Those two values can be considered good 
agreement, as the value of $\alpha$ must depend on the sample preparation 
through the strength of the non-magnetic impurity scattering as the constant
$\alpha$ depends exponentially on the mean free path (see Eq. 40b in Ref.\cite{label20})
as the size of the ballistic region. As it was pointed out by Fomin {\it et al.}
\cite{label21} the strength of the anisotropy is influenced also by surface
roughness, thus no better agreement could be expected. The value of $\alpha$ may be somewhat modified by taking into account the logarithmic correction.

The structure of level splitting depends on whether the impurity spin is 
integer or half-integer. In most of the cases the integer case $S=2$ have been
studied until now except the study Cu(Mn) alloys \cite{label10,label22}
where $S=5/2$ and the size dependence is essentially reduced. For completeness
such studies in magnetoresistance could be interesting from both experimental and theoretical point
of view.

In the calculation presented the logarithmic Kondo contribution is neglected, as the case $T\ll T_K$ is considered. Also the RKKY interaction between the impurities is neglected (see for discussion Ref.\cite{label1} and \cite{label23}). That can be justified by the high temperature, and the large level splitting considered also depresses their effect. The calculation is valid for samples with large mean free path. In the opposite limit the strength of the anisotropy is depressed and also the localization corrections can be dominant. For that limit the theory presented in Ref.\cite{label18} must be extended.

Finally we can conclude that Giordano's \cite{label1} magnetoresistance 
measurement of films with two different thickness provides an independent
support for existence of the proposed surface anisotropy.

\section*{Acknowledgement}
The authors are grateful for useful discussions with N. Giordano and O. \'Ujs\'aghy. The work was
supported by the Hungarian Grants Nos. OTKA 96 T021228 and T024005.
One of us (A.Z.) is grateful for the support by the Humboldt Foundation and 
the hospitality by the Meissner Institute of Low Temperature Physics and the 
Department of Physics of Ludwig Maximilians University in Munich.

\appendix
\section*{Eigenvalues of anisotropy Hamiltonian in case $S=2$ 
in a magnetic field ${\bf B}=(B,0,0)$ and for $S=1$ in arbitrary magnetic field.}
\label{app:1}

The eigenvalues of Hamiltonian (10) for $S=2$ are the following
\begin{eqnarray}
E_1&=&{5\over{2}} K_d\hbar ^2+\sqrt{9K_d^2\hbar ^4+4\mu ^2g^2\hbar ^2B^2}
\nonumber \\
E_2&=&{5\over{2}} K_d\hbar ^2-\sqrt{9K_d^2\hbar ^4+4\mu ^2g^2\hbar ^2B^2}
\nonumber \\
E_3&=&{5\over{3}} K_d\hbar ^2-{2\over{3}} \sqrt{13K_d^2\hbar ^4+12\mu ^2g^2\hbar
^2B^2} \nonumber \\
&\times &sin\left[
{{1\over 3}
arcsin\left(
{K_d\hbar ^2{{35K_d^2\hbar ^4
-72\mu ^2g^2\hbar ^2B^2}\over
{({13K_d^2\hbar ^4+12\mu ^2g^2\hbar ^2B^2})^{3\over 2}}}}
\right)}
\right]
\nonumber \\
E_4&=&{5\over{3}} K_d\hbar ^2+{2\over{3}} \sqrt{13K_d^2\hbar ^4+12\mu ^2g^2\hbar
^2B^2} \nonumber \\
&\times &sin\left[
{{1\over 3}
arcsin\left(
{K_d\hbar ^2{{35K_d^2\hbar ^4
-72\mu ^2g^2\hbar ^2B^2}\over
{({13K_d^2\hbar ^4+12\mu ^2g^2\hbar ^2B^2})^{3\over 2}}}}
\right)+{\pi\over 3}}
\right]
\nonumber \\
E_5&=&{5\over{3}} K_d\hbar ^2-{2\over{3}} \sqrt{13K_d^2\hbar ^4+12\mu ^2g^2\hbar
^2B^2} \nonumber \\
&\times &cos\left[
{{1\over 3}
arcsin\left(
{K_d\hbar ^2{{35K_d^2\hbar ^4
-72\mu ^2g^2\hbar ^2B^2}\over
{({13K_d^2\hbar ^4+12\mu ^2g^2\hbar ^2B^2})^{3\over 2}}}}
\right)+{\pi\over 6}}
\right],
\label{evbxs2}
\end{eqnarray}
which are shown in Fig. \ref{fig1}.

For $S=1$ and for arbitrary magnetic field the eigenvalues of the anisotropy Hamiltonian are
\begin{eqnarray}
E_1&=&{2\over{3}} K_d\hbar ^2+{2\over{3}}\sqrt{K_d^2\hbar ^4+3\mu ^2g^2\hbar ^2
(B_x^2+
B_z^2)}\nonumber\\
&\times & sin\left[
{{1\over 3}arcsin\left(
{K_d\hbar ^2{{K_d^2\hbar ^4+{9\over 2}\mu ^2
g^2\hbar ^2(B_x^2+2B_z^2)}\over {{(K_d^2\hbar ^4+3\mu ^2g^2\hbar ^2(B_x^2+
B_z^2))}^{3\over 2}}}}
\right)}
\right]
\nonumber \\
E_2&=&{2\over{3}}K_d\hbar ^2-{2\over{3}}\sqrt{K_d^2\hbar ^4+3\mu ^2g^2\hbar ^2(B_x^2+
B_z^2)}\nonumber\\
&\times &sin\left[
{{1\over 3}arcsin\left(
{K_d\hbar ^2{{K_d^2\hbar ^4+{9\over 2}\mu ^2
g^2\hbar ^2(B_x^2+2B_z^2)}\over {{(K_d^2\hbar ^4+3\mu ^2g^2\hbar ^2(B_x^2+
B_z^2))}^{3\over 2}}}}
\right)
+{{\pi}\over{3}}}
\right]
\nonumber \\
E_3&=&{2\over{3}}K_d\hbar ^2+{2\over{3}}\sqrt{K_d^2\hbar ^4+3\mu ^2g^2\hbar ^2(B_x^2+
B_z^2)}\nonumber\\
&\times &cos\left[
{{1\over 3}arcsin\left(
{K_d\hbar ^2{{K_d^2\hbar ^4+{9\over 2}\mu ^2
g^2\hbar ^2(B_x^2+2B_z^2)}\over {{(K_d^2\hbar ^4+3\mu ^2g^2\hbar ^2(B_x^2+
B_z^2))}^{3\over 2}}}}
\right)
+{{\pi}\over{6}}}
\right] ,
\label{evbzxs1}
\end{eqnarray}
which can be seen in Fig. \ref{fig2}.

\begin{figure}
\caption{The splitting of the spin eigenvalues is shown in magnetic field perpendicular to the surface without (dashed lines) and with (solid lines) surface anisotropy. The temperature scales $k_BT$ are placed at those magnetic fields where the lowest levels are separated from the others on larger scales than the temperature $k_BT$ (saturation fields).}
\label{fig0}
\end{figure}

\begin{figure}
\caption{Eigenvalues of Hamiltonian for impurity spin S=2 in magnetic field
parallel to the surface}
\label{fig1}
\end{figure}

\begin{figure}
\caption{Eigenvalues of Hamiltonian for impurity spin $S=1$ in
magnetic field
perpendicular to the surface (dashed line) and in case of $B_x/B_z=0.1$. 
(solid line) }
\label{fig2}
\end{figure}

\begin{figure}
\caption{The coordinate systems used for calculation of the eigenvalues 
and the magnetoresistance.}
\label{fig3}
\end{figure}

\begin{figure}
\caption{The conductivity vs magnetic field for a fixed $K_0$, in the case where the
magnetic field is perpendicular to the surface and $S=2$.}
\label{fig5}
\end{figure}

\begin{figure}
\caption{The conductivity vs magnetic field averaged over the positions of
the impurities, for magnetic field perpendicular to the surface, and $S=2$.
The calculation is carried out by adding of the results of 20 stripes with
anisotropy strengths given by Eq. (33).}
\label{fig6}
\end{figure}

\begin{figure}
\caption{The calculated conductivity vs magnetic field curves for magnetic
field perpendicular and parallel to the surface, for impurity spin S=1.}
\label{fig4}
\end{figure}

\begin{figure}
\caption{The magnetoresistance curves fitting the measurements of Giordano [1]
for Au(Fe) films of thickness 410 \AA { }  and 625 \AA { }. }
\label{fig7}
\end{figure}

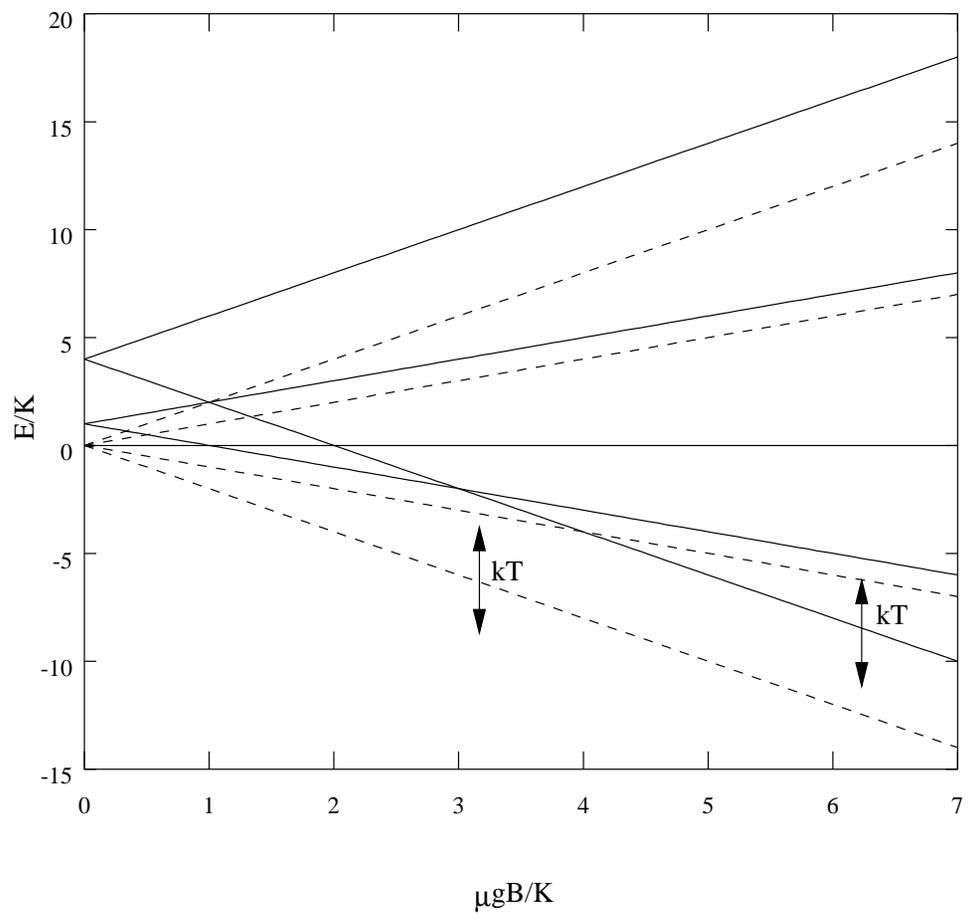
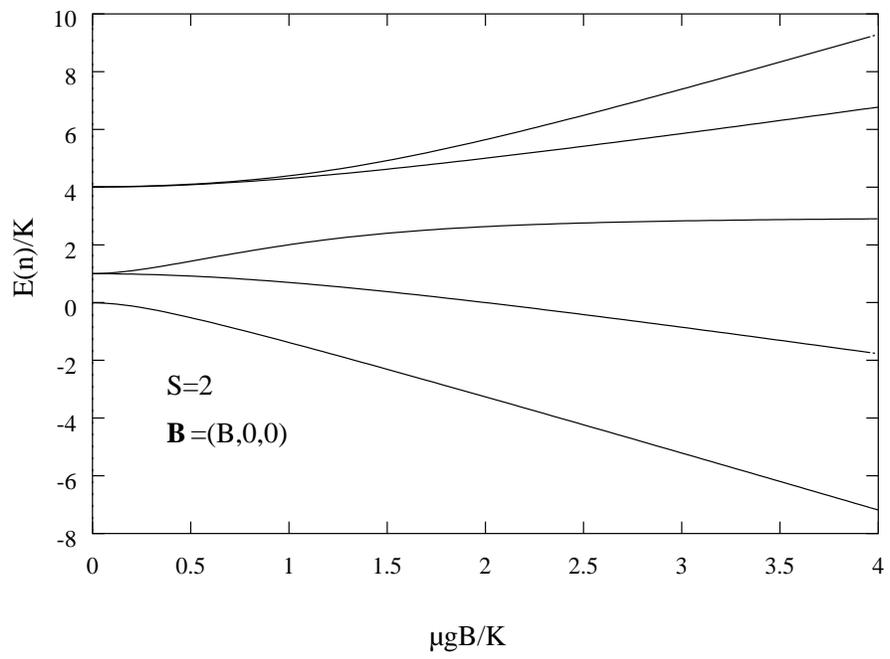
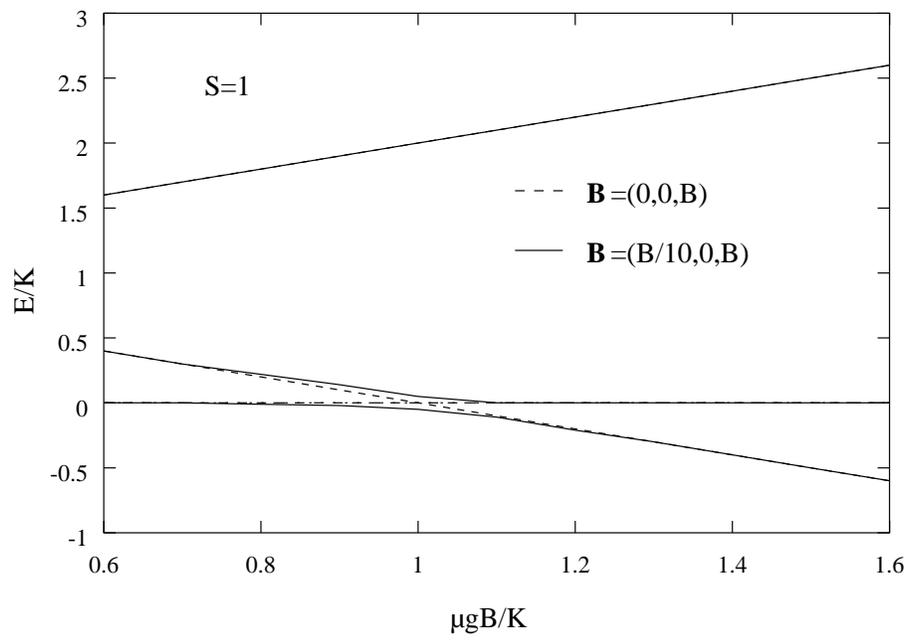
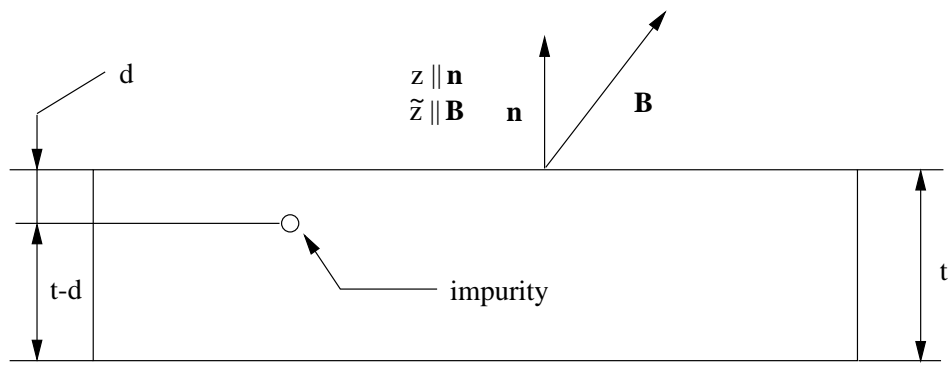
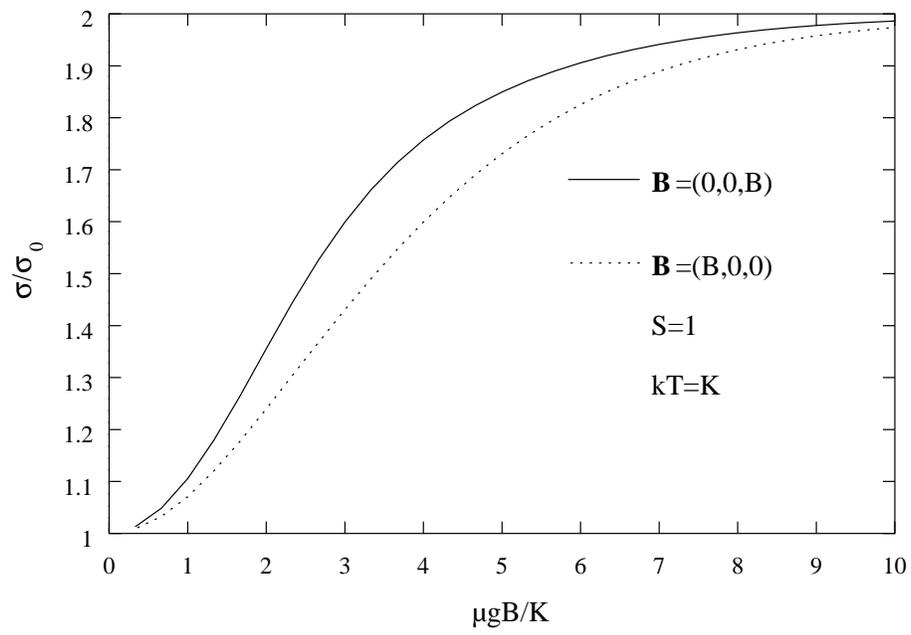
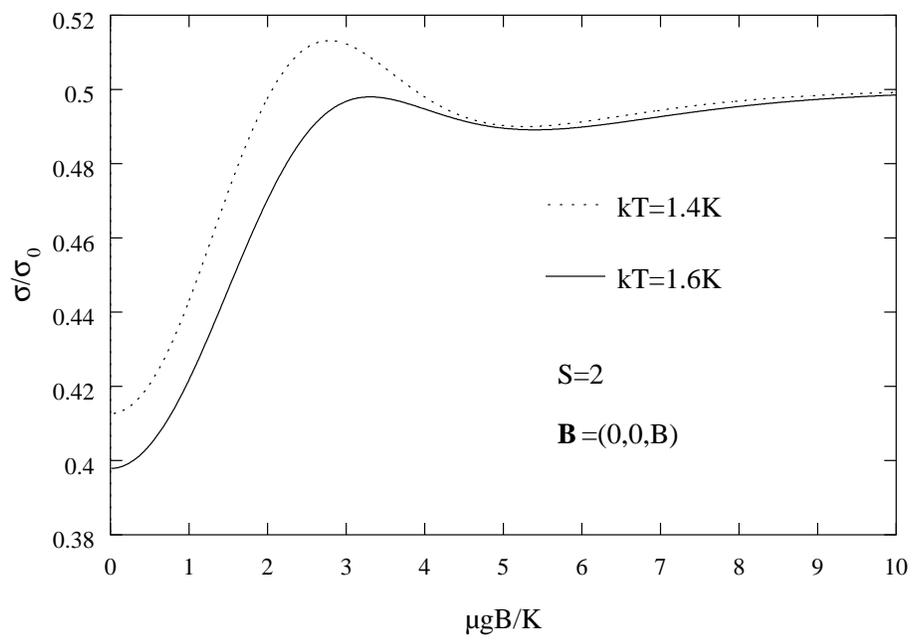
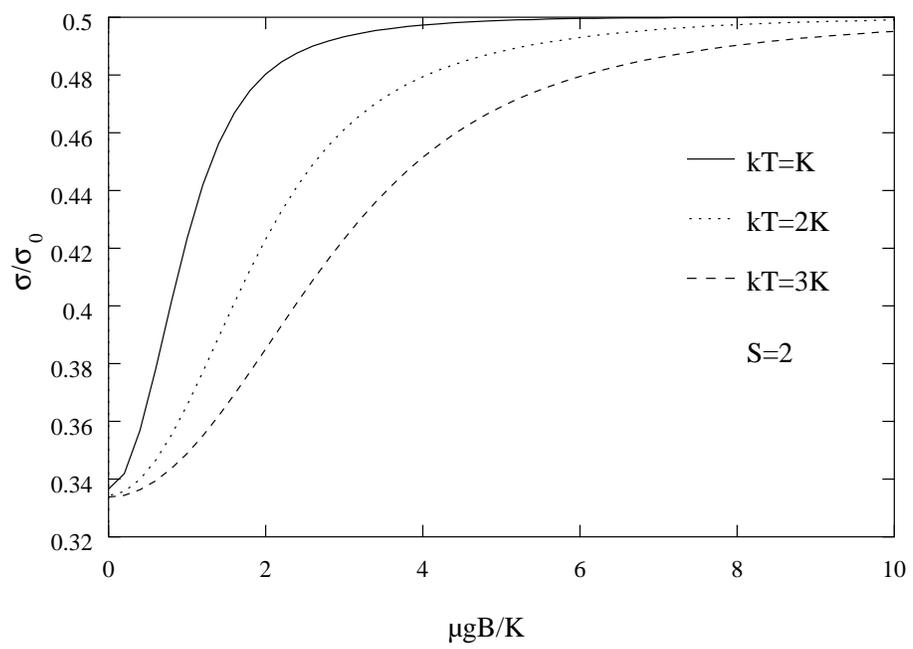
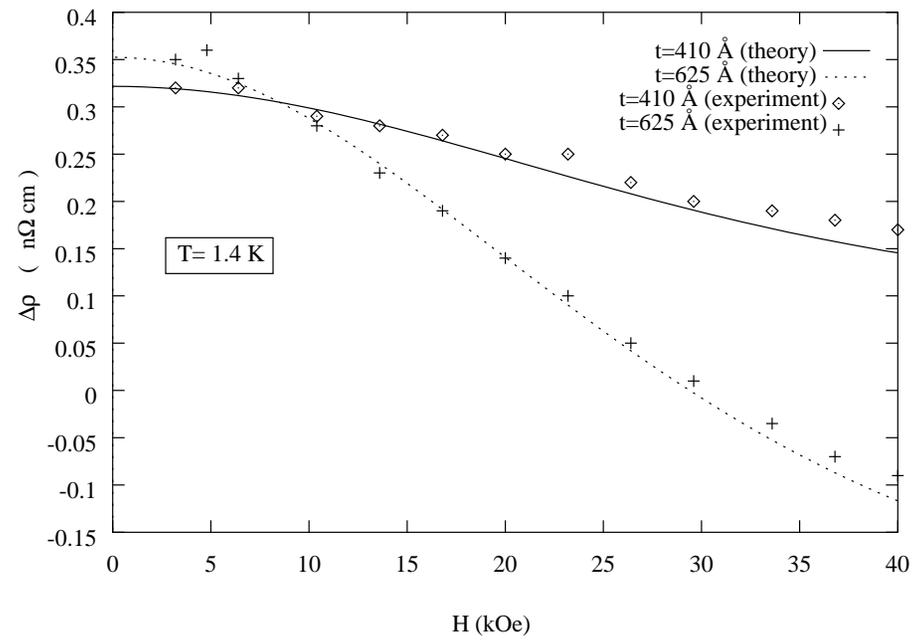
\end{document}